\documentclass[aps,pra,twocolumn,groupedaddress]{revtex4-1}
\usepackage[normalem]{ulem}
\usepackage{graphicx}
\usepackage{epstopdf}
\usepackage{amssymb}
\usepackage{url}
\begin{document}

\title{Polarization control of high-harmonic generation via the
spin-orbit interaction}


\author{Jack Wragg}
\email[]{jack.wragg@qub.ac.uk}
\author{Daniel D. A. Clarke}
\author{Gregory S. J. Armstrong}
\author{Andrew C. Brown}
\author{Connor P. Ballance}
\author{Hugo W. van der Hart}
\affiliation{Centre for Theoretical Atomic Molecular and Optical Physics, School of Mathematics and Physics, Queen's University Belfast, Belfast, Northern Ireland, BT7 1NN}


\date{\today}

\begin{abstract}
  We observe the generation of high harmonics in the plane perpendicular to the
  driving laser polarization and show that these are driven by the spin-orbit
  interaction. Using R-Matrix with time-dependence theory, we demonstrate that
  for certain initial states either circularly- or linearly-polarized
  harmonics arise via well-known selection rules between atomic states
  controlled by the spin-orbit interaction. Finally, we elucidate the
  connection between the observed harmonics and the phase of the intial state.
\end{abstract}

\pacs{}

\maketitle

Much progress in attosecond physics has been the result of advances in High
Harmonic Generation (HHG) techniques \cite{Schafer2018}. Enhanced understanding
and control of HHG has enabled ultrashort light sources \cite{43as_pulse},  and
myriad metrological techniques-- under the umbrella of `high-harmonic
spectroscopy'-- for the observation and manipulation of ultrafast electron
dynamics \cite{attosecond_spectroscopy_review}.

The HHG mechanism is well described by the `three-step model': the tunnel
ionization, acceleration and recombination of a valence electron driven by a
strong laser field resulting in the emission of photons of odd, integer
multiples of the driving laser photon energy \cite{Corkum1993}. The power of HHG
as a spectroscopic tool is that the attosecond-timescale dynamics of the target
atom are encoded in the spectrum of emitted light, and the rapid evolution of
both experimental techniques and underlying theory has resulted in a parade of
impressively sensitive and sophisticated measurements
\cite{Masin2018,Lai2018,Popruzhenko2018,Beaulieu2016,Kanai2005,Kanai2005}.

With this trend in mind, we may turn our attention beyond the
non-relativistic picture of electron dynamics, and seek to address the role of
the spin-orbit (SO) interaction in HHG. The SO interaction has two major effects
on the dynamics: (i) a splitting of the atomic states and (ii) a coupling of
these magnetic sublevels.
In principle, high-harmonic spectroscopic techniques should reveal all the
electronic processes which occur on the attosecond time-scale but, to date,
signatures of processes driven by the SO interaction have proved elusive.
The only effect reported in the literature is a small shift in the HHG cut-off
energy, caused by the the SO
splitting of the valence hole in krypton \cite{Pabst2014}. 
More recent studies have investigated the time-dependence of the SO interaction
in strong-field processes other than HHG
\cite{Hartung2016,Jordan2017}. Given the drive to manipulate electron dynamics
on the one hand, and the range of techiniques already available in high-harmonic
spectroscopy on the other, there is a clear motivation to address SO in HHG.

From a theoretical standpoint, the calculation of harmonic spectra is
straightforward: the harmonic spectrum is determined entirely from the
time-dependent expectation value of the dipole operator $\mathbf{D}$

\begin{equation}
  \mathbf{d}(t) = \langle{\Psi(t)|\mathbf{D}|\Psi(t)\rangle}
, 
\label{harms}
\end{equation}
where $\Psi(t)$ is the
wavefunction. However,
determining this quantity is only as straightfoward as the complexity of the
atomic-structure permits, and to include the SO interaction requires a highly
sophisticated theoretical approach. In this work we employ the R-matrix with
time-dependence (RMT) \cite{Moore2011,Hassouneh2014,Brown2013,Brown2016,Hamilton2017, Clarke2018, RMTurl} method to demonstrate that the SO
interaction leads to non-zero values of the $x$ and $y$ components of
$\mathbf{d}(t)$ when the driving laser pulse is polarized in the $z$ direction. 
This means that the presence of the SO interaction may be indicated by the
polarization of the HHG.

\begin{figure}
\includegraphics[width=0.45\textwidth]{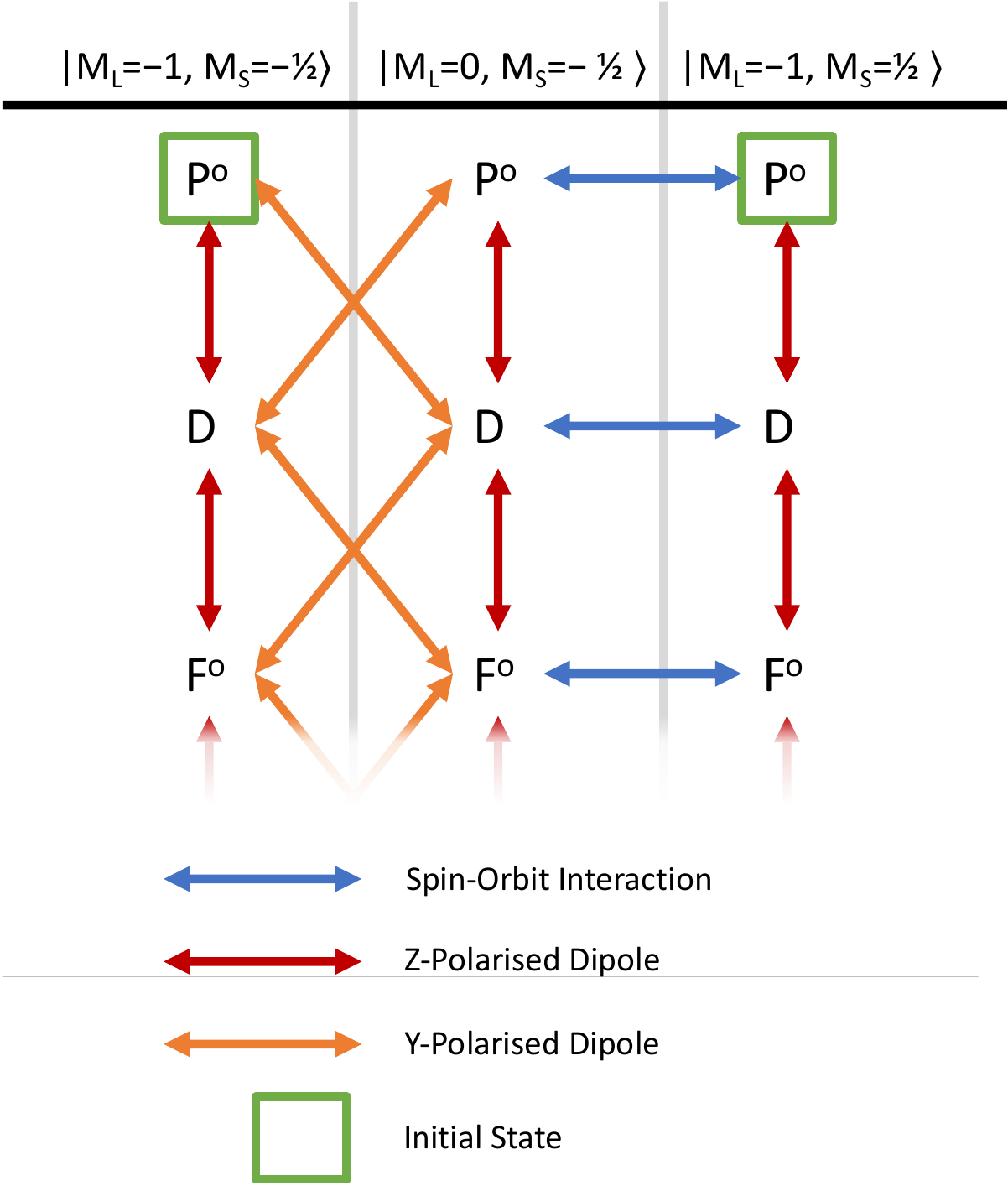}
\caption{Schematic demonstrating the process by which $y$-Polarized harmonic spectra can be produced through a combination of the spin-orbit interaction and a $z$-polarized driving field. This is a simplified description, and we omit the $S$ level for the $M_{L}=0$ states, and completely omit the $M_{L}=2$ states and associated $y$-polarized dipole interactions.}
\label{Demonstration}
\end{figure}

In order to simplify the atomic structure considerations, we employ a system
with a single, $^1S$ threshold: C$^+$, and we adopt, initially, the $LS$-coupling
framework for our discussion.
We will show that a C$^{+}$ ion, initially in some specific superposition of
states, will produce high-harmonics of a polarization
orthogonal to that of the driving field. Furthermore, we can show unambiguously 
that these harmonics arise due to the SO interaction.

In Fig. \ref{Demonstration}, we demonstrate the mechanism for these SO-induced
harmonics for an example ground state ($M_L=-1$) C$^{+}$ ion 
in a superposition of spin up and spin down.  Those states
which are coupled by the $y$ component of $\mathbf{D}$ must obey the
selection rules $\Delta M_{L}=\pm1$ and $\Delta M_{S}=0$. However, the
$z$-polarized driving field cannot change either the $M_{L}$ or $M_{S}$ of the
system, and thus the relevant states are not accessible. Thus $\mathbf{d}(t)$ only contains a component in the \(z\) direction. 

The action of the SO-interaction, however, is to transfer 
population from the $M_{L}=-1$ spin up states into the $M_{L}=0$
spin down states. These latter states {\it {are}} coupled by $\mathbf{D}_y$ to the $M_L = -1$ spin down states as the dipole selection rules
are satisfied. As a result, $\mathbf{d}_y(t)$ is non-zero, and harmonics are emitted polarized in the $y$ direction.

The generation of $y$-polarized harmonics can also be explained from the
$jK$-coupling perspective. This particular initial state is a
superposition of the $J=(3/2)^{o}$ 
(with contributions from both $M_{J}=1/2$ and $M_{J}=3/2$) 
and $J=(1/2)^{o}$ 
($M_{J}=1/2$) states. 
This means that
after a single transition (for example from $J=(3/2)^{o}$ to $J=(1/2)^o$) the $\Delta
M_{J}=\pm1$ selection rule for $\mathbf{D}_{y}$ is immediately satisfied,
and will remain so as the $z$-polarized driving pulse will not change
$M_{J}$. Without the SO-interaction, the radial
aspect the wavefunction corresponding to each $M_{J}$ value will propagate in
such a way that $\mathbf{D}_{y}$ in the positive and negative directions will cancel. As
such, unless the SO-interaction induces differences in the radial aspect
of the wavefunction, $\mathbf{d}_{y}(t)$ will still be zero, and harmonic
emission will be confined to $z$-polarization.




To test this hypothesis, we use RMT to solve the Schr\"odinger equation
incorporating the SO interaction
\begin{equation}
    \frac{d}{dt}\Psi(\mathbf{X},t)=-i\left[\hat{H}_{A}+\hat{H}_{SO}+\mathbf{E}(t)\cdot\mathbf{D}
      \right]\Psi(\mathbf{X},t), \label{tdse}
\end{equation}
where $\hat{H}_{A}$ is the time-independent, non-relativistic atomic
Hamiltonian, $\hat{H}_{SO}$ is the SO term from the Breit-Pauli
equation, and $\mathbf{E}(t)$
is the time-dependent electric field strength of the driving laser pulse, polarized in the \(z\) direction. From
our chosen initial wavefunction, we then obtain the wavefunction at
subsequent times solving Eq. (\ref{tdse}) iteratively. At each time step we calculate the expectation value
of $\mathbf{D}$ using Eq. (\ref{harms}). We then obtain the harmonic spectrum of the
radiation emitted by the atom polarized in the $x$, $y$ and $z$ directions from the
Fourier transforms of 
the $x, y$, and $z$ components of $\mathbf{d}(t)$ respectively.

For these calculations we consider a C$^+$ ion in two initial superpositions,
described in Tabs. \ref{TableTwo} and \ref{SupA}. The ion description is built
up by coupling an electron to bound orbitals for a C$^{2+}$ target state
\cite{Clementi1974}. These continuum orbitals are constructed from a basis of 50
B-Splines of 9th order. For all results, the laser pulse is an 8 cycle 800nm
pulse (3 cycles ramp on, 2 cycles at peak intensity, 3 cycles ramp off) of peak
intensity 10$^{14}$ W/cm$^{2}$. The field is linearly polarized in the $z$
direction and propagates in the $x$ direction.

\begin{figure}
\includegraphics[width=0.45\textwidth]{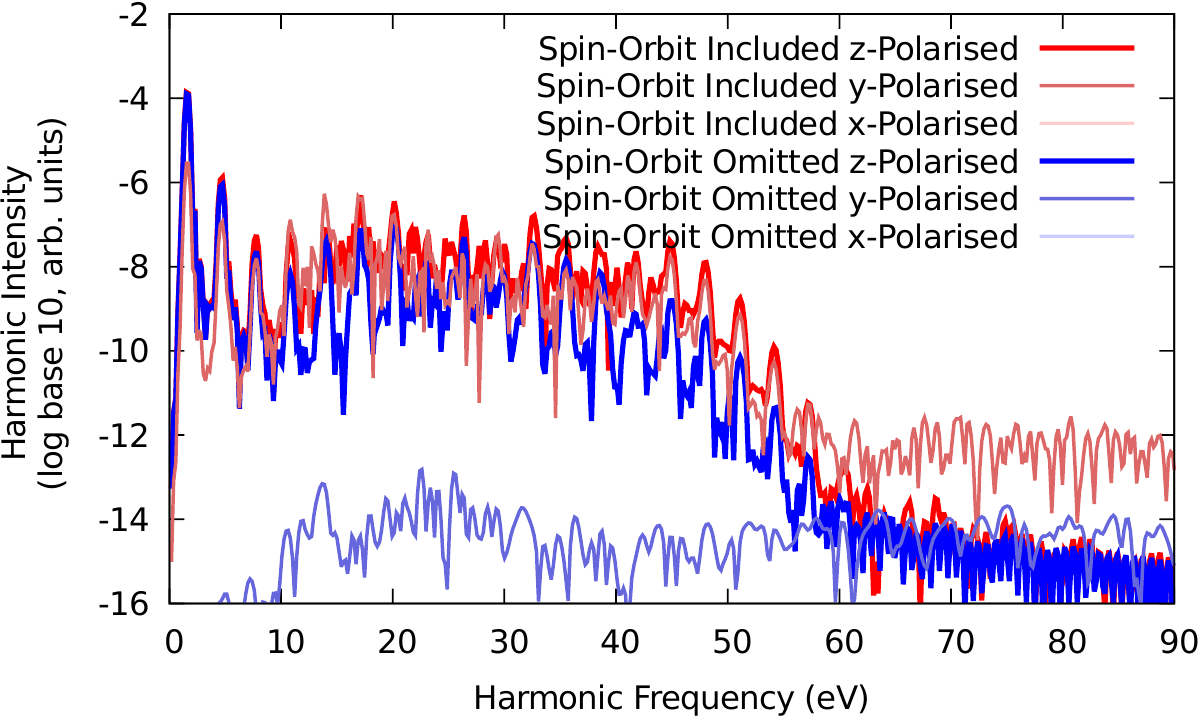}
\caption{HHG spectra from a C$^{+}$ ion in an initial superposition given in table \ref{TableTwo} in an 800nm laser pulse of 10$^{14}$ W/cm$^{2}$. We show spectra for cases were the spin-orbit interaction is included (Red) and omitted (Blue). For each case we then show the $z$ polarized component (darker line) and the $x$ and $y$ components (lighter lines)}.
\label{PlotTwo}
\end{figure}

Figure \ref{PlotTwo} shows the HHG spectrum calculated with and without the
SO-interaction. The intial state is as described in Tab. \ref{TableTwo}-- a
superposition of $M_L=-1$ spin up and $M_L=1$ spin down-- where we
have adopted an initial phase difference of $\theta =0$. As expected, omitting the SO-interaction yields harmonics
polarized only in the $z$-direction, whereas including the SO-interaction gives
harmonics with both $y$ and $z$ polarization. 


\begin{table}
\centering
\begin{tabular}{ c |  c | c | c }
 L &  $M_{L}$ & $M_{S}$ & Weight \\ 
\hline  $1$ & $-1$ & $\frac{1}{2}$ & $\sqrt{\frac{1}{2}}$ \\  
  $1$ & $1$ & $-\frac{1}{2}$ & $e^{i\theta}\sqrt{\frac{1}{2}}$    
\end{tabular}
\hspace{1cm}
\begin{tabular}{  c | c | c }
   J & $M_{J}$ & Weight \\ 
\hline  $\frac{1}{2}$ & $-\frac{1}{2}$ & $-\sqrt{\frac{1}{3}}$\\  
  $\frac{3}{2}$ & $-\frac{1}{2}$ & $\sqrt{\frac{1}{6}}$\\  
  $\frac{1}{2}$ & $\frac{1}{2}$ & $e^{i\theta}\sqrt{\frac{1}{3}}$\\  
  $\frac{3}{2}$ & $\frac{1}{2}$ & $e^{i\theta}\sqrt{\frac{1}{6}}$\\  
\end{tabular} 
\caption{The initial-state composition used for figures \ref{PlotTwo} and
\ref{Plot18}, in both LS (left) and jK (right) coupling. $\theta$ is the phase difference in the initial superposition.}
\label{TableTwo}
\end{table}

\begin{figure}
\includegraphics[width=0.45\textwidth]{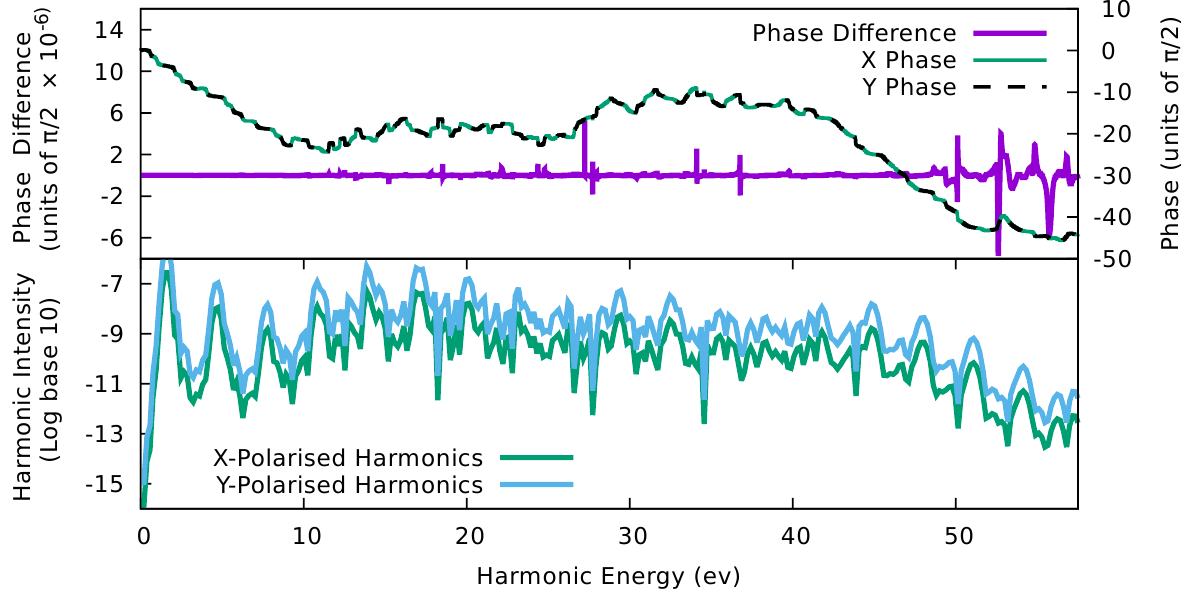}
\caption{Harmonic intensity, phase, and relative phase between the $x$ and $y$ components of the spectrum obtained from a C$^{+}$ ion in an initial superposition in table \ref{TableTwo} (in this case, $\theta=\pi/10$). The harmonic phase and relative phase between the $x$ and $y$ harmonics (upper pane) is shown over a wider range than the usual $-\pi...\pi$ to more clearly show the continuous nature of the data. The intensity is plotted over the same energy region for the same harmonic spectrum (lower pane)}.
\label{Plot18}
\end{figure}

Further calculations reveal that the polarization of the harmonic spectrum in
the $x/y$ plane depends on the phase-difference, $\theta$, in the initial-state
superposition. Specifically, we find that the $x$ and $y$ components of the
harmonic spectrum combine to produce a linearly polarized harmonic spectrum of
angle $\theta$ off the $y$ axis. Figure
\ref{Plot18} shows the results from an initial state with a phase difference of $\theta=\pi/10$ 
within the superposition. Both
$x$ and $y$ polarized harmonics are produced, although $y$ polarized harmonics
are still of higher intensity. Moreover, the phases of the $x$ and $y$ polarized
harmonics are identical throughout the spectrum, indicating that the harmonics
are again linearly polarized. The angle from the $y$-axis is calculated from
the intensity of the individual $x$ and $y$ polarized harmonics, is found
to be equal to the phase difference in the initial superposition,
$\theta$. This relationship between $\theta$ and the angle of polarization in
the $x$-$y$ plane was found to hold true across angles in the range $\theta=0
\ldots \pi/2$.


We now consider the superposition described in table \ref{SupA}, where the
C$^{+}$ ion is in the $M_{L}=-1$ state with a superposition of spin up and spin
down. This means that after the effect of the SO interaction there will
be a $y$-polarized dipole between the $M_{L}=-2$ and $M_{L}=-1$ spin up states,
and another between the $M_{L}=-1$ and $M_{L}=0$ spin down states.  Again, we
add an initial phase factor $e^{i\theta}$ into the spin-down component. 

\begin{table}
\centering
\begin{tabular}{ c | c | c | c }
$L$ &  $M_{L}$ & $M_{S}$ & Weight \\ 
\hline   $1$ & $-1$ & $\frac{1}{2}$ & $\sqrt{\frac{1}{2}}$ \\  
 $1$ & $-1$ & $-\frac{1}{2}$ & $e^{i\theta}\sqrt{\frac{1}{2}}$    
\end{tabular}
\hspace{1cm}
\begin{tabular}{  c | c | c }
   J & $M_{J}$ & Weight \\ 
\hline  $\frac{1}{2}$ & $-\frac{1}{2}$ & $-\sqrt{\frac{1}{3}}$\\  
  $\frac{3}{2}$ & $-\frac{1}{2}$ & $\sqrt{\frac{1}{6}}$\\  
  $\frac{3}{2}$ & $-\frac{3}{2}$ & $e^{i\theta}\sqrt{\frac{1}{2}}$\\  
\end{tabular} 
\caption{The initial state composition used for figures \ref{PlotOne} and
\ref{PhasePlot} in both LS (left) and jK (right) coupling. $\theta$ is the phase difference in the initial superposition.}
\label{SupA}
\end{table}

\begin{figure}
\includegraphics[width=0.45\textwidth]{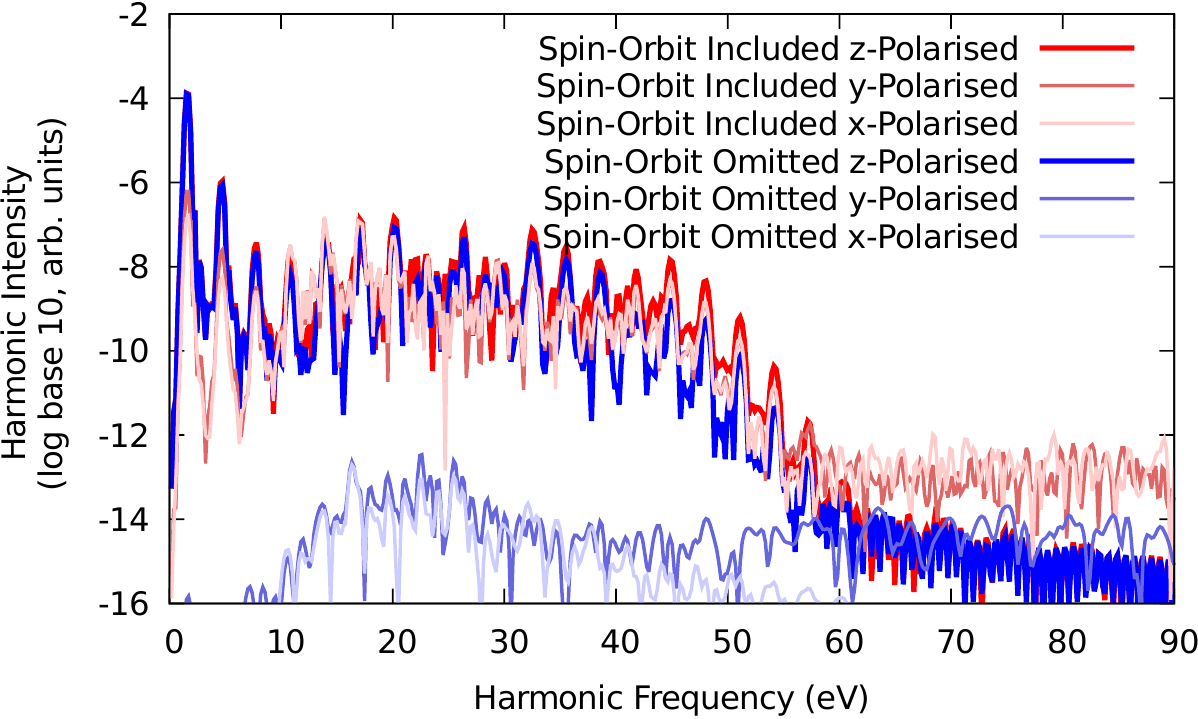}
\caption{HHG spectra from a C$^{+}$ ion in an initial superposition given in table \ref{SupA} in an 800nm laser pulse of 10$^{14}$ W/cm$^{2}$. We show spectra for cases were the spin-orbit interaction is included (Red) and omitted (Blue). For each case we then show the $z$ polarized component (darker line) and the $x$ and $y$ components (lighter lines)}.
\label{PlotOne}
\end{figure}

\begin{figure}
\includegraphics[width=0.45\textwidth]{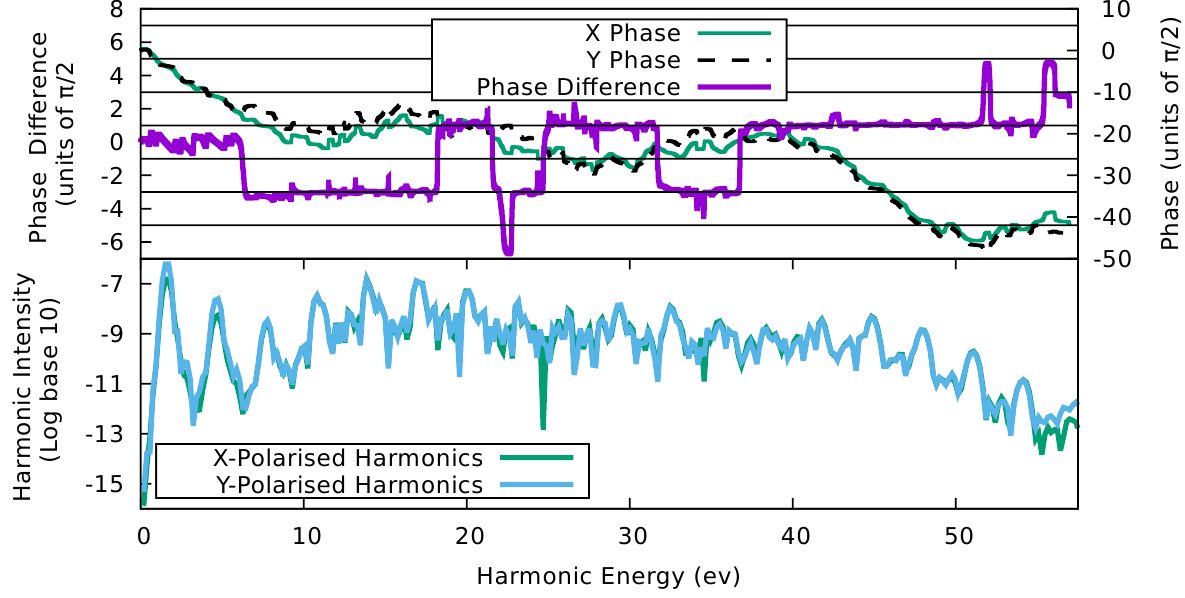}
\caption{Harmonic intensity, phase, and relative phase between the $x$ and $y$ components of the spectrum obtained from a C$^{+}$ ion in an initial superposition in table \ref{SupA}. The harmonic phase and relative phase between the $x$ and $y$ harmonics (upper pane) is shown over a wider range than the usual $-\pi...\pi$ to more clearly show the continuous nature of the data. The intensity is plotted over the same energy region for the same harmonic spectrum (lower pane)}.
\label{PhasePlot}
\end{figure}

Figure \ref{PlotOne} shows the $x$, $y$ and $z$ components of the HHG spectrum.
As before, both $y$ and $z$ polarized harmonics are produced, but unlike the
previous initial superposition, here $x$-polarized harmonics are also present.

Fig. \ref{PhasePlot} shows that the $x$ and $y$ harmonic spectra have an
identical amplitude throughout most of the energy range. The phase difference
between the $x$ and $y$ polarized harmonics persists at a half-integer multiple
of $\pi$, indicating circularly polarized harmonics (as all harmonics are of the
same intensity). We note that there is no clear pattern in the
phase-relationship in the $x$/$z$ or $y$/$z$ planes. 
Finally, we find that for initial states of $\theta\ne 0$, there is no significant change in the character of the harmonic spectra.  


There has been much interest recently in extending HHG to the generation of
elliptical and circular harmonics. This is typically attempted using two-color
counter-rotating circular pulses \cite{long1995}, or pulses of varying linear
polarization \cite{neufeld2017}.  Here it is an atomic mechanism-- the
SO interaction-- rather than the character of the laser field, that
provides a means of generating multi-dimensional harmonics.  
Although we have restricted our results to C$^+$, these results should persist
in general. In those systems with a doublet ground state consisting of
a single $p$-hole, such as noble gas ions, we would expect $y$-polarized
HHG via the same mechanism described in Fig. \ref{Demonstration}.

In this work we have considered only the single atom response to the
driving field, and as such we have included a description of the resulting $x$
polarized spectra. However, as the driving field propagation is in the $x$
direction, observing the $x$ polarized harmonic spectra will be challenging.
Nonetheless, these results show interesting
observations are possible using only the $z$- and $y$-polarized harmonic spectra. For example, 
we could use the intensity of the emitted $y$-polarized harmonics as a measurement of the phase difference between two
states in a superposition. This phase difference will most often have
arisen as a result of the methods used to create the initial superposition,
however when the phase difference is unknown this could be a useful experimental
tool.

The question remains of how to produce such initial states. 
A linearly polarized pulse will preferentially eject an electron of $m_{\ell}=0$.
It might be suggested therefore, that a short pulse might be used to eject such
an electron from a noble gas atom to form a target of a superposition of
$M_{L}=0$ spin up and spin down. We have suggested a different approach 
in \cite{Wragg2019} where the $m_{\ell}$ of the ejected electron can be
controlled by dipole selection rules when exciting an individual electron to the
only energetically accessible $s$ state. This method has the advantage of having
been demonstrated theoretically using ultrashort laser pulses, providing 
greater of control over the purity of the initial superposition. 



In conclusion, we have demonstrated that given an appropriate initial state, the
presence of the spin-orbit interaction will cause harmonic spectra to be
generated of a polarization orthogonal to that of the driving field.
The specific superposition chosen may lead to linear or circular
polarization in the $x$/$y$ plane. As such, these results provide a clear route
to observe spin-orbit effects through HHG processes

\begin{acknowledgments}
We acknowledge Jakub Benda and Zden\v ek Ma\v s\' in for their collaboration in developing and maintaining the RMT code. The data presented in this article may be accessed at Ref. \cite{DataDoi}. The RMT code is part of the UK-AMOR suite, and can be obtained for free at Ref. \cite{RMTurl}. This work benefited from computational support by CoSeC, the Computational Science Centre for Research Communities, through CCPQ. D.D.A.C. acknowledges financial support from the UK Engineering and Physical Sciences Research Council (EPSRC). A.C.B., H.W.v.d.H., G.S.J.A and J.W. acknowledge funding from the EPSRC under Grants No. EP/P022146/1, No. EP/P013953/1, and No. EP/R029342/1. This work relied on the ARCHER UK National Supercomputing Service \cite{Archerurl}, for which access was obtained via the UK-AMOR consortium funded by EPSRC.
\end{acknowledgments}

\bibliography{Main}

\end{document}